\documentclass[conference]{IEEEtran}
\IEEEoverridecommandlockouts

\usepackage{amsmath,amssymb,amsfonts}
\usepackage{algorithmic}
\usepackage{graphicx}
\usepackage{textcomp}
\usepackage{xcolor}

\def\BibTeX{{\rm B\kern-.05em{\sc i\kern-.025em b}\kern-.08em
    T\kern-.1667em\lower.7ex\hbox{E}\kern-.125emX}}

\begin{document}

\title{On SSI-based Private Decentralized Bidding\\
}

\author{\IEEEauthorblockN{1\textsuperscript{st} Andreea-Elena Dr\u agnoiu}
\IEEEauthorblockA{\textit{Department of Computer Science} \\
\textit{University of Bucharest}\\
\textit{Research Institute of the University of} \\
\textit{Bucharest (ICUB)}\\
Bucharest, Romania \\
andreea-elena.panait@drd.unibuc.ro}
\and
\IEEEauthorblockN{2\textsuperscript{nd} Nicoleta Dumitru}
\IEEEauthorblockA{\textit{Department of Mathematics} \\
\textit{University of Bucharest}\\
\textit{Research Institute of the University of} \\
\textit{Bucharest (ICUB)}\\
Bucharest, Romania \\
nicoleta.dumitru@my.fmi.unibuc.ro}
\and
\IEEEauthorblockN{3\textsuperscript{rd} Ruxandra F. Olimid}
\IEEEauthorblockA{\textit{Department of Computer Science} \\
\textit{University of Bucharest}\\
\textit{Research Institute of the University of} \\
\textit{Bucharest (ICUB)}\\
Bucharest, Romania \\
ruxandra.olimid@fmi.unibuc.ro}
}

\maketitle

\begin{abstract}
In private bidding, participants submit sealed bids that remain hidden (also from other bidders) during the bidding window. This is essential in competitive environments to ensure a fair and independent evaluation of proposals. While (public) blockchain enables decentralized bidding, and its transparency offers advantages such as public verifiability, without a Trusted Third Party (TTP), current methods struggle to verify whether a bidder is eligible. We propose a framework that addresses this issue by certifying bid eligibility and authenticity using the established Self-Sovereign Identity (SSI) framework. To let participants prove that they meet the necessary bidding requirements while keeping their bids secret, our model uses Verifiable Credentials (VCs) and cryptographic primitives such as Zero-Knowledge Proofs (ZKPs). 
\end{abstract}


\section{Introduction}
\label{sec:introduction}

Although bidding and auction are sometimes used interchangeably in everyday language, there is a clear distinction between them in a technical context. An auction is a dynamic, competitive, and interactive process that allows for the continued placement of bids, followed by the response of the other bidders over the duration of the auction, in a number of rounds. In contrast to the one-time submission characteristic of bidding, auctions involve real-time feedback and sequential price adjustments, culminating in the selection of a winner at the end of the process. 

This study focuses exclusively on bidding because its linear, formally structured nature is especially suitable for blockchain-based applications. The decentralized architecture of blockchain enables the creation of a secure and immutable record of each bid, ensuring that submissions cannot be altered once recorded. In addition, smart contracts can automate the evaluation and selection process according to predefined rules. As a result, the procedure becomes transparent, verifiable, and independent of centralized control, thereby reducing the need for trust among participants. Ultimately, this approach strengthens procedural integrity and ensures that outcomes are determined solely by objective, pre-programmed criteria.

\subsection{Motivation}

The classical, decentralized commit-reveal scheme on blockchain, while an improvement over centralized bidding, still has a major weakness: the blockchain is transparent by default, meaning that participants' identities (generally anonymized) and, once revealed, bid values become permanently public. In competitive settings, this creates real risks (a losing bidder's strategy, capabilities, and valuation are exposed to all future competitors). More critically, the system lacks a mechanism to verify who is actually bidding. Anyone can submit a commitment, meaning eligibility, reputation, and authorization cannot be enforced without reintroducing a Trusted Third Party (TTP).
Using Self-Sovereign Identity (SSI), participants hold and control their own verifiable credentials, selectively disclosing only what is necessary to prove eligibility without revealing anything beyond what the policy requires.

\subsection{Novelty and Contributions}

This study considers bidding as a decentralized mechanism for selecting among competing proposals in the context of SSI. We model participants as self-sovereign agents who submit sealed bids derived from their own Verifiable Credentials (VCs) and locally controlled data in response to a predefined request. These submissions are evaluated according to fixed criteria, and a single winner is determined without any iterative interaction or negotiation among participants. This sealed-bid setting is particularly relevant in privacy-sensitive and adversarial environments, as it ensures that each participant's strategy is based solely on privately held information, capabilities, and cryptographically verifiable claims, rather than observing or reacting to other bidders.

To the best of our knowledge, we are the first to propose incorporating the established SSI framework into private bidding and thus making use of its advantages in these settings.
Our work brings the following main contributions:

\begin{enumerate}
    \item We consider SSI-based private bidding and define a general framework.
    \item We compare our proposed framework to the existing general framework for private bidding.
    \item We explicitly look into various properties (classified on several classes) and their satisfiability within the two frameworks.
\end{enumerate}

\subsection{Outline}

The remainder of the paper is organized as follows. Section~\ref{sec_relw} discusses related work. Section \ref{sec_ssi} introduces SSI.
Section~\ref{sec_framework} presents the classical, general framework for private bidding using the blockchain and the proposed enhanced framework that leverages SSI functionality. Section \ref{sec_properties} compares the properties of the two frameworks. Section \ref{sec_discussion} briefly discusses our proposal. 
Section~\ref{sec_conclusion} concludes.

\section{Related work}
\label{sec_relw}

Several blockchain-based bidding and auction protocols have been proposed over the years. However, none combines private bidding with SSI capabilities. Kosba et al.~\cite{Kosba2016HawkTB} used Multi-Party Computation (MPC) but with high interaction costs, while Yuan et al.~\cite{Yuan2018} used TEE, which requires hardware updates and has security concerns~\cite{davenport14, aumasson16}. Other works rely on semi-trusted auctioneers \cite{BlaKer18}, suffer from inefficiency and lack fund verification on Ethereum \cite{GalYou18}, cannot support sealed-bid auctions \cite{NguTha21}, or allow bidder collusion \cite{QusTarAkr20}.
In \cite{chen2022building}, the authors consider private bidding and user identity protection in case of coercion. 
In \cite{luzhang21}, the authors introduce a blockchain-based sealed-bid domain name auction protocol that uses smart contracts, the Pedersen commitment, and Zero-Knowledge Proofs (ZKPs). The domain name bidder can bid in a Vickrey-type auction that lets the bidder keep the bid secret until all bids are submitted, and the winning bidder pays the second-highest bid instead of his bid~\cite{luzhang21}. 
A privacy-preserving sealed-bid auction protocol using TEE execution and on-chain settlement \cite{Gebele25} was demonstrated on Ethereum and SUAVE. It achieves confidentiality and verifiable results without trusted intermediaries or protocol changes. Despite some residual trust and typical TEE security risks, the protocol shows that practical private auctions for DeFi applications are feasible today \cite{Gebele25}.
In \cite{zhang24}, a Homomorphic Encryption (HE)-based bid comparison circuit was designed to determine the winning bid in a sealed-bid auction while allowing bidders to verify the computation without revealing their bids. Based on this circuit, a Blockchain-based Sealed-bid Scheme (BSS) using commitments and zero-knowledge proofs was proposed. The scheme was formally proven secure against bid leakage and winner manipulation, and its performance is evaluated to assess efficiency and practicality.
\cite{Dai24} presents a blockchain-based electronic bidding auction system that protects privacy and resists malicious participants. It uses secure MPC and threshold elliptic curve cryptography to eliminate the need for a TTP. The protocol is proven secure, and simulations show is is efficient and practical.
In \cite{bouaicha25}, the authors propose a blockchain-based framework that prevents online auctions from shill bidding behaviour through dynamic, behaviour-based penalties enforced by smart contracts. Using a Bid Shill Score (BSS) to evaluate nine bidding patterns, the system adjusts penalties to make fraudulent activity economically unprofitable while ensuring fair competition.
The study \cite{Kokaras23} analyzes privacy-preserving cryptographic mechanisms for blockchain platforms, especially for Ethereum, and proposes an anonymous auction protocol based on Anon-Zether \cite{Diamond21}. Implemented through a smart contract in Solidity, the protocol hides both bidders’ identities and bid amounts while an auctioneer determines the result. Although this higher anonymity increases gas costs compared to some sealed-bid auctions, it remains cheaper than fully decentralized approaches without an auctioneer.
The paper \cite{Chin22} presents a sealed-bid auction protocol that hides both the bid value and the maximum bid while ensuring bidders have sufficient funds (fund binding). It uses indistinguishable Ethereum transactions and the DECO protocol \cite{deco20} to prove fund availability without revealing sensitive information, and evaluates the implementation against deposit-based auction methods.
In \cite{ChenLiXiangWang22}, the authors propose a decentralized, smart contract–based sealed-bid reverse auction that claims to preserve bid privacy, resist collusion, and require no trusted third party. Using ZKPs and anonymous veto networks, it ensures security and verifiability, with efficient performance demonstrated on an Ethereum-based prototype.
\cite{Tyagi23} introduces practical, fully decentralized sealed-bid auction protocols using timed commitments. These ensure fairness even with dropouts or collusion, while keeping bids hidden. The scheme includes non-malleable commitments with range proofs to verify sufficient funds, enabling penalties for abandoned bids and support for multiple auctions. An Ethereum-based implementation shows the approach is efficient and practical.
\cite{Chen24} proposes BERA, a blockchain-based system for securely allocating edge computing resources. It combines a sealed-bid auction using Ethereum smart contracts and privacy-preserving techniques with a graph neural network to detect fraud without revealing bids.
In \cite{Vangujar24}, the authors propose BCCIBI, a blockchain-based scheme to improve e-auction security and privacy. It combines multiple cryptographic techniques, such as ElGamal encryption, BLS signatures, and ZKPs, to ensure bid confidentiality, bidder anonymity, and secure verification.

\section{Self-Sovereign Identity (SSI)}
\label{sec_ssi}

Self-Sovereign Identity (SSI) is a digital identity model that empowers individuals with control over their identities. This framework uses a decentralized approach to enable privacy-preserving, context-dependent (personal) data sharing, which significantly mitigates the risks of identity theft and reduces reliance on centralized third parties \cite{MuhHruGayMei18}, \cite{DunPet18}, \cite{DragnoiuOlimid2024}. The SSI ecosystem functions through a structured three-party model involving distinct roles and digital artifacts~\cite{MuhHruGayMei18}, \cite{DragnoiuCiobanuOlimid2025}. The \textbf{holder}, typically the user, owns and manages their identity data within a digital wallet. The \textbf{issuer}, typically a recognized authority, creates and cryptographically signs assertions about the holder. To complete the transaction, the \textbf{verifier} validates these assertions by checking the proof against a Verifiable Data Registry (VDR)—often a blockchain—to ensure the information is authentic and has not been revoked~\cite{DragnoiuOlimid2024,w3c_verifcred}.

The data exchanged within this ecosystem takes the form of Verifiable Credentials (VCs) and Verifiable Presentations (VPs). A VC is a digital, tamper-evident proof of specific claims that includes cryptographic metadata, and it is cryptographically signed by the issuer~\cite{DragnoiuOlimid2024,w3c_verifcred}. VCs are typically stored off-chain to maintain user privacy and are used to generate a VP (which can incorporate a single VC or several VCs), which is the actual data shared with a verifier. This mechanism allows the holder to present only the minimal information for a specific context, thus enabling selective disclosure and ensuring that private data remains secure while still being verifiable across different platforms~\cite{DragnoiuOlimid2024, DragnoiuCiobanuOlimid2025,w3c_verifcred,AhmIslShaIsl22,tessaro2023revisiting}.

\section{Private Decentralized Bidding Frameworks}
\label{sec_framework}

We further discuss the general private bidding framework that uses blockchain, as already presented in the literature \cite{bouaicha25,ChenLiXiangWang22,luzhang21,LiXue21}. Then, we propose an SSI-based framework that brings in the advantages of self-sovereignty and makes the best use of the established SSI framework.
 
\subsection{General Private Bidding (on Blockchain)}

The general framework involves three primary roles: a \textit{requestor}, a group of \textit{bidders}, and a \textit{smart contract} that governs their interactions on a public blockchain.
Note that the terminology is not fully settled in the literature; for example, \cite{ChenLiXiangWang22} refers to an \textit{auctioneer} and multiple \textit{bidders}, \cite{luzhang21} refers to a \textit{domain name owner} and \textit{domain name bidders}, and \cite{bouaicha25} refers to a \textit{seller} who owns the item, multiple \textit{bidders}, and an \textit{auctioneer} who conducts the bidding procedure.

The main reason for using a public blockchain is to simplify the bidding process by allowing any bidder the possibility to place a bid, without any approval or prior interaction with the requestor that runs the bidding, while also making the overall process transparent and publicly verifiable.

\textbf{Requestor.} The requestor is the entity that runs the bidding, and it is directly interested in the result. 
The requestor initiates the process by deploying a smart contract that encodes all bidding logic, eligibility rules, evaluation criteria, and timing constraints. Once deployed, this contract is immutable and executes autonomously. The requestor has no further privileged control over the outcome.

\textbf{Smart contract.} The contract acts as the neutral coordinator of the entire process. It enforces the sequencing of phases, validates incoming submissions, timestamps, and stores all data on-chain, and ultimately determines and publishes the winner according to its predefined logic.

\textbf{Bidders.} Each bidder participates independently, placing its own bid. This is performed by executing the smart contract. 
Bidders have no visibility into each other's submission values at any point in the process.
Nevertheless, smart contract execution is generally public (we normally consider public blockchains), exposing information such as the number of placed bids.

A bid can contain one or more values, depending on the requirements. The winning bid is the one that meets the requirements best, with the \textit{clearing price} being the price that the winning bidders pay.

\begin{figure}[t!]
    \centerline{\includegraphics[width=\columnwidth]{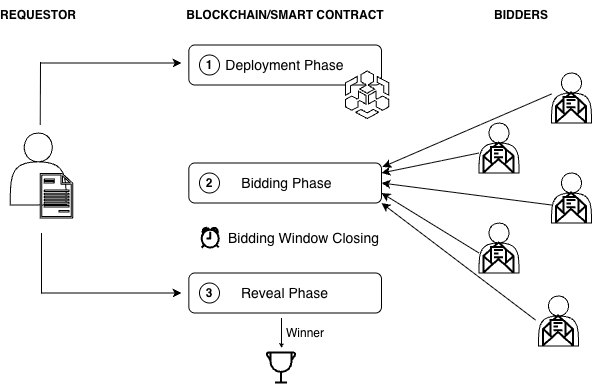}}
    \caption{General bidding on blockchain. The protocol proceeds in three phases: (1) \textit{Deployment Phase} - the requestor deploys the bidding logic; (2) \textit{Bidding Phase} - bidders submit cryptographic hiding and bidding bids (commitments) 
    during the open window; (3) \textit{Reveal Phase} - bidders reveal their actual values after the window closes, enabling the smart contract to verify, evaluate, and announce the winner without the necessity of a TTP.}
    \label{fig:generalModel}
\end{figure}

\smallskip
The process unfolds in three sequential phases, as depicted in Figure \ref{fig:generalModel}.

\textbf{1. Deployment Phase.} The requestor publishes the smart contract, establishing the rules and opening the \textit{bidding window}, a predefined time period during which bidders can submit their bids.

\textbf{2. Bidding Phase.} Each bidder places a bid by accessing the smart contract, typically using a cryptographic commitment (e.g., a hash of their bid value combined with a secret nonce). 
Because of the commitment's properties, the committed value does not reveal the bid value (the \textit{hiding} property) and prevents changes (the \textit{binding} property).
The contract records each submission with a timestamp, ensuring integrity and immutability.

\textbf{3. Reveal Phase.} After the bidding window closes, the bidders disclose their original bid values and nonces.  
The smart contract layer verifies the commitments, and only bids that pass this verification are accepted for evaluation.
This step is necessary because the blockchain's inherent transparency would otherwise allow late bidders to observe and undercut earlier submissions.  
The smart contract layer then applies its predefined selection and announces the winner on-chain. 
We leave out of the discussion here how exactly the bidders open their commitments; a straightforward approach is for bidders to open their commitments publicly, and for the requestor to trigger an \textit{end-auction} transaction on the smart contract to finalize the process.

\subsection{SSI-based Private Bidding (on Blockchain)}

\begin{figure*}[h]
    \centerline{\includegraphics[width=.8\textwidth]{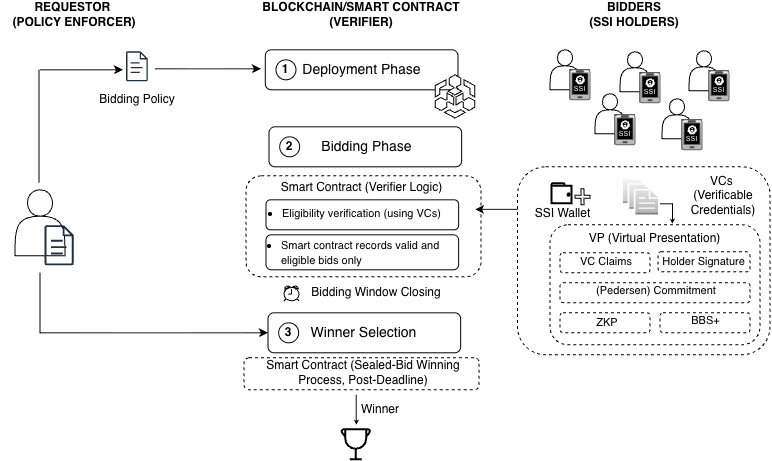}} 
    \caption{SSI-based private bidding on blockchain, an SSI-augmented commit-reveal framework in which eligibility is enforced cryptographically rather than by the necessity of a TTP. The protocol proceeds in three phases: (1) \textit{Deployment Phase} - the requestor deploys the bidding logic, including an eligibility policy in the smart contract; (2) \textit{Bidding Phase} - during the open window, bidders submit cryptographic commitments as well as eligibility and identity proofs using SSI capabilities (VPs); (3) \textit{Winner Selection} - after the bidding window closes, the contract validates all proofs, evaluates the sealed bids, and publishes the winner on-chain, preserving both bidder privacy and result verifiability. }
    
    \label{fig:SSIModel}
\end{figure*}

In SSI-based settings, bidders hold VCs issued by trusted authorities that attest to relevant claims, such as identity, certifications, financial capacity, or eligibility requirements. These credentials allow bidders to participate without relying on a centralized authority to validate them during the bidding process.

Similar to the general framework, we maintain the same actors and unfold the process in three sequential phases depicted in Figure \ref{fig:SSIModel}.

\textbf{1. Deployment Phase.} It maintains the same goal as in the classical framework, but explicitly enforces, at the verifier (smart contract) level, a policy specifying which claims a bidder must prove to demonstrate eligibility to submit an offer. 

\textbf{2. Bidding Phase.} This phase extends the general framework by requiring bidders to explicitly prove eligibility under the enforced policy before placing a bid. For clarity, we treat \textit{Eligibility verification}, \textit{Bid submission}, and \textit{Bid validation and storage} as three separate sub-steps; however, the first two could naturally be merged.

\textbf{Eligibility verification}. 
Before submitting an offer, the bidder must prove eligibility. For this, the bidder constructs a VP from one or more VCs.
By using special cryptographic constructions, such as the BBS+ signature scheme \cite{tessaro2023revisiting}, the verifier learns only the necessary information while all other attributes remain hidden, enforcing \textit{selective disclosure}. The bidder signs the VP and submits it to the verifier, which validates the credentials and confirms eligibility. Only eligible bidders are allowed to proceed.

\textbf{Bid submission}. Once eligibility is confirmed, the bidder submits the actual bid. As in the general framework, a commitment that enforces both the hiding and bidding properties (such as the Pedersen commitment) could be used. 
Additionally, a ZKP can be attached to prove that the committed bid satisfies predefined auction rules (e.g., valid range or format) without revealing the underlying data.

\textbf{Bid validation and storage}. The smart contract records only valid bid commitments. 
This way, the blockchain does not store sensitive bidder information or plain text bid values, but only cryptographic proofs or commitments that guarantee correctness and fairness.

\textbf{3. Winner Selection}. During the final phase, which is similar to the \textit{Reveal phase}, the winner is selected. 
Only the winner’s identifier (possibly anonymized) and the clearing price are published. To achieve this, one can enforce privacy-preserving mechanisms to determine the winner.

\medskip
Unlike classical bidding schemes, where smart contracts mainly act as record keepers, the SSI-based model introduces active verification of bidder eligibility before bids are recorded and makes direct use of certification of attributes in the form of VCs/VPs.

\smallskip
\textbf{Adversarial model.} The adversarial threat is naturally defined in bidding systems, with a focus on the bidders. While the smart contract itself can be a point of attack, the requestor is considered honest. The reason is that dishonest requesters could directly enforce eligibility or winning criteria to their advantage anyway. A weak adversary is a passive, honest-but-curious entity seeking to disclose restricted information from the bidding process, such as the real identities of the bidders, the final ranking, or the bid values (also during the bidding phase). A stronger active, malicious adversary that attempts to place bids adaptively to maximize winning bids or to impersonate another party by stealing or cloning VPs should be considered. Coalitions of malicious parties are naturally defined.

\section{Properties}
\label{sec_properties}

We further investigate properties of interest for private decentralized bidding.
Table \ref{tab:1} summarizes the results, describing these properties and indicating whether they are satisfied in both the general private bidding framework and the SSI-based private bidding framework. We classified the properties into the following categories: \textit{Privacy \& Data Protection}, \textit{Identity}, \textit{Trust \& Governance}, \textit{Security \& Performance}, and \textit{Correctness \& Verifiability}.

\begin{table*}[ht]
\centering
\caption{Comparison of Properties: General Framework vs. SSI-based Framework} 
\label{tab:1}
\begin{tabular}{p{0.15\textwidth}p{0.13\textwidth} p{0.40\textwidth}p{0.1\textwidth}p{0.1\textwidth}}
\hline
\textbf{Property Name} & \textbf{Type} & \textbf{Description} & \textbf{General Framework} & \textbf{SSI-based Framework}\\
\hline

Privacy of losing bids \cite{Vangujar24} &  Privacy \& Data Protection & Only the winning bid is disclosed, the bids submitted by participants who do not win the auction remain confidential and are not revealed to others & Yes & Yes\\ 

Anonymity \cite{DragnoiuCiobanuOlimid2025,LiXue21} & Privacy \& Data Protection & The identities of the bidders are kept hidden during the bidding process & Yes & Yes\\

Unlinkability \cite{DragnoiuCiobanuOlimid2025} & Privacy \& Data Protection & Individual bids cannot be linked to the identity of the bidder or to other bids made by the same bidder. & Partially & Yes\\

Data Ownership \cite{MuhHruGayMei18,DragnoiuCiobanuOlimid2025}& Privacy \& Data Protection & Bidders retain control and rights over the data they submit during bidding & Partially & Yes\\ 

Selective Disclosure \cite{DragnoiuOlimid2024, DragnoiuCiobanuOlimid2025,AhmIslShaIsl22,tessaro2023revisiting} & Privacy \& Data Protection & A bidder reveals only the specific credential attributes required to qualify for or participate in a bid & No & Yes \\

\hline

Self-Sovereignty \cite{MuhHruGayMei18,DragnoiuOlimid2024,DragnoiuCiobanuOlimid2025}& Identity, Trust \& Governance & Bidders have full ownership and control over their personal data, identity, or assets, and decide with whom to share the data & Partially & Yes\\

Issuer Credential Independence \cite{DunPet18,DragnoiuOlimid2024,DragnoiuCiobanuOlimid2025} & Identity, Trust \& Governance & Holders can present VPs without being dependent on or traceable back to the original issuer during the bidding process & No & Yes\\

Credential Portability \cite{Vangujar24,DragnoiuCiobanuOlimid2025,AhmIslShaIsl22}& Identity, Trust \& Governance & VCs issued in different ecosystems can be presented and recognized across multiple, unrelated bidding environments & No & Yes\\

Identity verification \cite{Vangujar24, DragnoiuCiobanuOlimid2025} & Identity, Trust \& Governance & Verifying the identity of participants before they are allowed to take part in the bidding process & Yes (up to a TTP) & Yes\\  

Eligibility Check & Identity, Trust \& Governance & Confirm the eligibility of bidders to take part within the bidding process  & Yes (up to a TTP) & Yes \\

Overall trust \cite{MuhHruGayMei18,DunPet18,DragnoiuCiobanuOlimid2025}& Identity, Trust \& Governance & Confidence of the participants that the bidding process is honest, fair, and will be executed as promised & Partially & Yes\\ 

\hline

Aliveness \cite{Vangujar24} & Security \& Performance & A bidding system remains active, responsive, and capable of progressing toward a valid outcome, even in the presence of failures, uncooperative participants, or adversarial conditions & Yes & Yes\\ 

Scalability \cite{Vangujar24} & Security \& Performance & Maintain performance, correctness, and efficiency as the number of bidders, bids, or auction complexity grows & Yes & Yes\\ 

Decentralization \cite{chen2022building,DragnoiuOlimid2024,DragnoiuCiobanuOlimid2025,LiXue21} & Security \& Performance & Distribution of control, authority, and execution across multiple independent parties  & Yes (up to a TTP) & Yes \\ 

Fraud Detection \cite{Vangujar24} & Security \& Performance & Mechanisms used in bidding systems to identify and prevent dishonest or illegal activities during the bidding process & Yes & Yes\\ 

Non-repudiation \cite{MuhHruGayMei18, DragnoiuOlimid2024, DragnoiuCiobanuOlimid2025} & Accountability  & A bidder cannot later deny having submitted, agreed to, or authorized a bid & Partially & Yes\\

\hline

Correctness \cite{Vangujar24, DragnoiuCiobanuOlimid2025} & Correctness \& Verifiability & Ensures the bid is accurate, complete, and error-free before submission & Yes & Yes\\ 

Data Verifiability \cite{DragnoiuCiobanuOlimid2025}& Correctness \& Verifiability & Independently confirm the accuracy and authenticity of data used in bidding processes & Partially & Yes\\

Public verifiability/Fairness \cite{LiXue21} & Correctness \& Verifiability & The ability of anyone to independently verify that a bidding process was conducted correctly, without accessing confidential bid details & Partially & Yes\\

\hline
\end{tabular}
\end{table*}

\section{Discussion}
\label{sec_discussion}

Intuitively, general private bidding on the blockchain hides bid values and ensures a fair bidding process. However, its correct termination assumes honest bidders in the sense that they are allowed to participate in the bidding process and that they place correct bids.
For example, bidders should not be allowed to bid more than they can afford to pay. Nevertheless, there is no guarantee that the bidders have the funds to pay the clearance price. The SSI-based framework solves this by proving that the bidder owns the necessary funds, in the form of a VP (satisfying \textit{Data ownership} and \textit{Data verifiability} in Table \ref{tab:1}). Also, SSI-based identification fully maintains \textit{Decentralization} (also for identification and eligibility checks) while enforcing \textit{Self-Sovereignty}: the bidder is fully in charge of their own identity and uses selective disclosure to prove only the necessary attributes to establish their eligibility in the bidding process. Fully eliminating the need for a TTP increases overall \textit{Trust} in the system, and \textit{Selective disclosure} limits data exposure, thereby directly facilitating other properties, such as \textit{Unlinkability}. 
While incorporating SSI eliminates the need for a TTP during auction execution, trust is relocated to the VPs used to prove eligibility. This trade-off is advantageous because eligibility checks rely on established SSI mechanisms rather than interactive, centralized verification during the bidding process itself. This shows the superiority of the proposed framework in terms of privacy and data protection, security, correctness and verifiability, trust, and, naturally, identity management and self-sovereignty.

Moreover, the SSI framework is well established and standardized \cite{w3c_did, w3c_verifcred}, while the bidding process is not, lacking unified rules and suffering from platform fragmentation. Thus, using the SSI framework for bidding might facilitate similar adoption and provide a standardized method for certifying bids. This directly reduces the assumed trust level, thereby increasing confidence in an honest and fair bidding process.

Technically, an SSI-based bidding system can be implemented by integrating an SSI framework, blockchain infrastructure, and other cryptographic tools. The \textbf{SSI layer} relies on W3C standards \cite{w3c_did, w3c_verifcred}, and can be managed using a framework such as Hyperledger Aries, Hyperledger Indy, or Veramo. These frameworks are usually integrated in backend services developed in Node.js, Python, or Java, and should use digital wallets to allow bidders to store and present credentials.
The \textbf{execution layer} can be implemented on blockchain platforms, such as Ethereum or Hyperledger Fabric, with smart contracts being commonly written in Solidity for Ethereum-based systems, while Hyperledger Fabric uses chaincode written in Go, Java, or Node.js. 
The \textbf{cryptographic mechanisms} can be implemented using tools like Circom, snakjs, or ZoKrates. The \textbf{frontend layer} can use frameworks like React or Angular.
A prototype of the proposed solution is left for future work, along with other aspects such as the type of blockchain used (public vs. private) and extensions to multi-round auctions.

\section{\uppercase{Conclusions}}
\label{sec_conclusion}

We proposed an SSI-based bidding framework that incorporates SSI features into a general blockchain-based bidding system. We have investigated the advantages SSI brings in terms of privacy, security, verifiability, and trust, compared to the classical, general decentralized bidding framework. Our informal analysis indicates that the proposed model is superior: it provides stronger security guarantees, simplifies verifications such as identity and eligibility checks, preserves data privacy, and enables selective disclosure. Moreover, SSI brings in standardization, as SSI is now a well-established field and its adoption is clearly defined by current regulations. Our current work is pioneering SSI-based private decentralized bidding.

\section*{Acknowledgment}
This work was supported by a grant of the Ministry of Research, Innovation and Digitalization, CNCS/CCCDI - UEFISCDI, project number ERANET-CHISTERA-IV-PATTERN, within PNCDI IV.

\bibliographystyle{IEEEtran}
\bibliography{listofpubs}

\end{document}